\documentclass[10pt,twocolumn,letterpaper]{article}

 \usepackage[pagenumbers]{cvpr} 

\usepackage{url}
\usepackage{graphicx}
\usepackage{booktabs}
\usepackage{amsthm}
\newtheorem{proposition}{Proposition}

\usepackage{verbatim}
\usepackage{enumitem}

\usepackage{amsmath,amsfonts,bm}

\def\eqref#1{equation~\ref{#1}}

\def\1{\bm{1}}

\DeclareMathAlphabet{\mathsfit}{\encodingdefault}{\sfdefault}{m}{sl}
\SetMathAlphabet{\mathsfit}{bold}{\encodingdefault}{\sfdefault}{bx}{n}

\definecolor{cvprblue}{rgb}{0.21,0.49,0.74}
\usepackage[pagebackref,breaklinks,colorlinks,allcolors=cvprblue]{hyperref}

\def\confName{CVPR}
\def\confYear{2027}

\title{Equivariant Experience Replay for Continual Learning in Imaging Inverse Problems}

\author{
Guixian Xu \quad Jin Zhu \quad Jinglai Li \quad Junqi Tang*\\
School of Mathematics\\
University of Birmingham\\
{\tt\small
gxx422@student.bham.ac.uk,
\{j.zhu.7,
j.li.10,
j.tang.2\}@bham.ac.uk
}
}

\begin{document}
\maketitle
\begin{abstract}

In this work, we propose Equivariant Experience Replay, a new paradigm for achieving continual learning of deep networks for solving ill-posed imaging inverse problems. Inspired by recent advances on unsupervised learning methods such as equivariant imaging/splitting, we find the exploitation of group-symmetry structure is particularly relevant and effective in continual learning setting for ill-posed inversion tasks. We therefore propose this new experience replay scheme, which is tailored for such inverse problems. Our numerical study demonstrates the superiority of our method compared to existing continual learning methods for inverse problems in medical imaging and image processing.

\end{abstract}    
\section{Introduction}

Even in this era of deep learning and AI, real world imaging tasks, such as medical imaging inverse problems, are all operating under imperfect scenarios. For instance, measurements could be limited in number and particularly noisy, the computational resource for practitioners can be very limited, pretrained AI models may face distribution shifts in practice, just to name a few. There is an urgent need of developing a principled and tailored framework for continual learning in inverse problems under such highly imperfect scenarios in order to maximize the adaptivity of AI models in imaging practice.

Recent advances in deep learning for imaging inverse problems have focused on improving practical flexibility in aspects such as unsupervised training \citep{chen2021equivariant,sechaud2026equivariant} and test-time training \citep{darestani2022test, zhao2024test,xu2026equivariant}. Starting from the work by \cite{chen2021equivariant} on Equivariant Imaging (EI), researchers have been developing unsupervised training frameworks which are able to bypass the need for ground-truth data. Such learning paradigms enable effective learning of the null-space of the measurement operator, which is particularly crucial for successful training of deep networks in solving ill-posed inverse problems. Meanwhile, test-time training (TTT) techniques have also been developed recently to improve the robustness of inference in imaging inverse problems. TTT schemes are typically applied when the task at hand is outside the training distribution of the pretrained model. 

While current TTT schemes are particularly effective for mitigating distribution shifts and provide better reconstruction consistently via performing inference-time fine-tuning of deep networks, they cannot be directly extended to continual learning tasks due to catastrophic forgetting. In continual learning setting, we wish to maintain the accuracy of the deep models on the pretrained tasks, while at the same time excel on the new task which has different measurement physics and data distribution. Although we can observe that the generic continual learning schemes in literature, such as Experience Replay \citep{rolnick2019experience} and its variants \citep{buzzega2020dark}, can be extended to solving inverse problems, there is currently no tailored continual learning method for inverse problems which are often ill-posed. Building on recent successes in unsupervised training, in this work, we propose Equivariant Experience Replay (EER) which is the first tailored continual learning scheme for imaging inverse problems, and demonstrate that utilizing the group-symmetry structure for ill-posed measurement operators is particularly important in achieving high-quality continual learning.

\subsection{Related Works}

\textbf{Continual learning (CL).} CL studies how to sequentially update a model on a stream of tasks so that it adapts to new tasks while mitigating catastrophic forgetting of previously learned ones~\citep{wang2024comprehensive}. Existing methods broadly fall into three categories.
\begin{itemize}[leftmargin=*]
\item \emph{Memory-based methods} retain a small buffer of samples from previous tasks and replay them jointly with new-task data during learning new tasks~\citep{buzzega2020dark,farajtabar2020orthogonal,arani2022learning,sarfraz2023error,yoo2024layerwise,tong2025coreset,liu2026ider}. 
\item \emph{Regularization-based methods} penalize changes to parameters or explicitly restrict update directions that are important for previous tasks, thereby preserving the knowledge for old tasks~\citep{kirkpatrick2017overcoming,wang2021training,lin2022trgp,lin2023theory,magistri2024elastic,li2025theory,liu2026elastic}. 
\item \emph{Architecture-based methods} dynamically introduce new parameters, such as additional modules, tokens, prompts, or low-rank adapters, as new tasks arrive~\citep{chaudhry2018efficient,serra2018overcoming,yan2021dynamically,douillard2022dytox,smith2023coda,liang2024inflora,wu2025sdlora,le2026one}. 
\end{itemize}
Most CL methods target supervised classification, and CL in medical imaging has mainly addressed segmentation and detection~\citep{karani2018lifelong,baweja2018towards,hofmanninger2020dynamic,perkonigg2021dynamic}, both of which assume labelled data. Continual reconstruction differs in two respects: the forward operator shifts across tasks, and the replay buffer holds only measurements and their operators, without ground truth~\citep{knoll2020advancing,antun2020instabilities}. We therefore build on memory- and regularization-based CL but replace supervised replay with our self-supervised equivariant replay, which derives virtual operators from stored measurements to constrain the nullspace of the stored operators.

\textbf{Data-driven solutions for inverse problems}. In recent years, the computational imaging community has been focusing on developing data-driven algorithms for solving imaging inverse problems. By utilizing both measurement physics and deep learning-based priors, these data driven approaches are able to provide significantly improved accuracy comparing to classical model-driven approaches. Starting from the invention and development of plug-and-play priors \citep{venkatakrishnan2013plug,kamilov2017plug,tan2024provably}, researchers have massively advanced the frontier of deep learning for inverse problem, both in terms of specialized architectures \citep{ronneberger2015u,zhang2017beyond} and flexible training schemes, such as self-supervised / unsupervised training \citep{lehtinen2018noise2noise,chen2021equivariant,tan2024unsupervised,carioni2024unsupervised,sechaud2026equivariant}, and test-time training \citep{ulyanov2018deep, darestani2022test}. Going beyond unsupervised learning and test-time training, our work here initializes the research of principled continual learning tailored for inverse problems.
\section{Algorithmic Framework}\label{sec:method}

\subsection{Problem Formulation and Overview}\label{sec:formulation}

\paragraph{From continual learning to continual inverse problems.}
Standard continual learning (CL) methods are primarily designed for domain- or class-incremental settings and do not account for the physical measurement constraints of imaging inverse problems. In continual inverse problems, the target is an image $x$, and the input is a measurement produced by a \emph{known} physical acquisition process,
\begin{equation}
    y = A x + \epsilon, \qquad x \sim p(x), \quad A \in \mathcal{A},
    \label{eq:forward}
\end{equation}
where $A$ is the forward operator and $\epsilon$ is measurement noise. A task is therefore determined by three factors: the image distribution $p_i(x)$, the acquisition physics $\mathcal{A}_i$ (e.g., a distribution over undersampling masks), and the noise model $\nu_i$:
\begin{equation}
    \mathcal{T}_i = \big(p_i(x),\ \mathcal{A}_i ,\nu_i \big).
    \label{eq:task}
\end{equation}
 The acquisition physics can change between tasks, $\mathcal{A}_i \neq \mathcal{A}_{i+1}$, even when the image distribution stays the same, $p_i = p_{i+1}$. We refer to this change as \emph{operator shift}. Typical examples are a change of acceleration factor or sampling pattern in MRI and a change of the number or angular range of projections in CT. Such changes occur routinely within a single scanner as acquisition protocols are updated, and reconstruction networks trained under one sampling scheme are known to degrade under another~\citep{darestani2022test}. Operator shift is the focus of this work; Our formulation also allows simultaneous shifts in the image distribution, $p_i \neq p_{i+1}$, and in the noise model, $\nu_i \neq \nu_{i+1}$.


\paragraph{Learning protocol.}
The reconstruction network $\hat{x} = \mathcal{F}_{\theta}(y, A)$ is trained sequentially across tasks $\mathcal{T}_1,\ldots,\mathcal{T}_N$. At stage $i$, the learner observes only measurements from the current task, $\mathcal{D}_i = \{(y^{(k)}, A^{(k)}, \epsilon^{(k)})\}_{k=1}^{n_i}$, with no access to ground-truth images, together with a memory buffer $\mathcal{B}_{i-1} \subseteq \bigcup_{j<i} \mathcal{D}_j$ of fixed capacity $|\mathcal{B}_{i-1}| \le K$, where $K \ll \sum_{j<i} n_j$. Starting from $\theta_{i-1}$, the goal is to obtain parameters $\theta_i$ that perform well on all tasks seen so far:
\begin{equation}
    \theta_i \in \arg\min_{\theta}\ \sum_{j=1}^{i}\ \mathbb{E}_{x \sim p_j,\, A \sim \mathcal{A}_j,\, \epsilon}\Big[\ell\big(\mathcal{F}_{\theta}(Ax+\epsilon, A),\, x\big)\Big].
    \label{eq:objective}
\end{equation}
At test time the operator $A$ is given but the task index is not, which places our setting closest to domain-incremental learning~\citep{van2022three}. Since $x$ is never observed, \eqref{eq:objective} cannot be accessed directly and must be approximated with self-supervised losses built from $\mathcal{D}_i$ and $\mathcal{B}_{i-1}$. 

\begin{figure*}[t]
    \centering
    \includegraphics[width=1.00\textwidth]{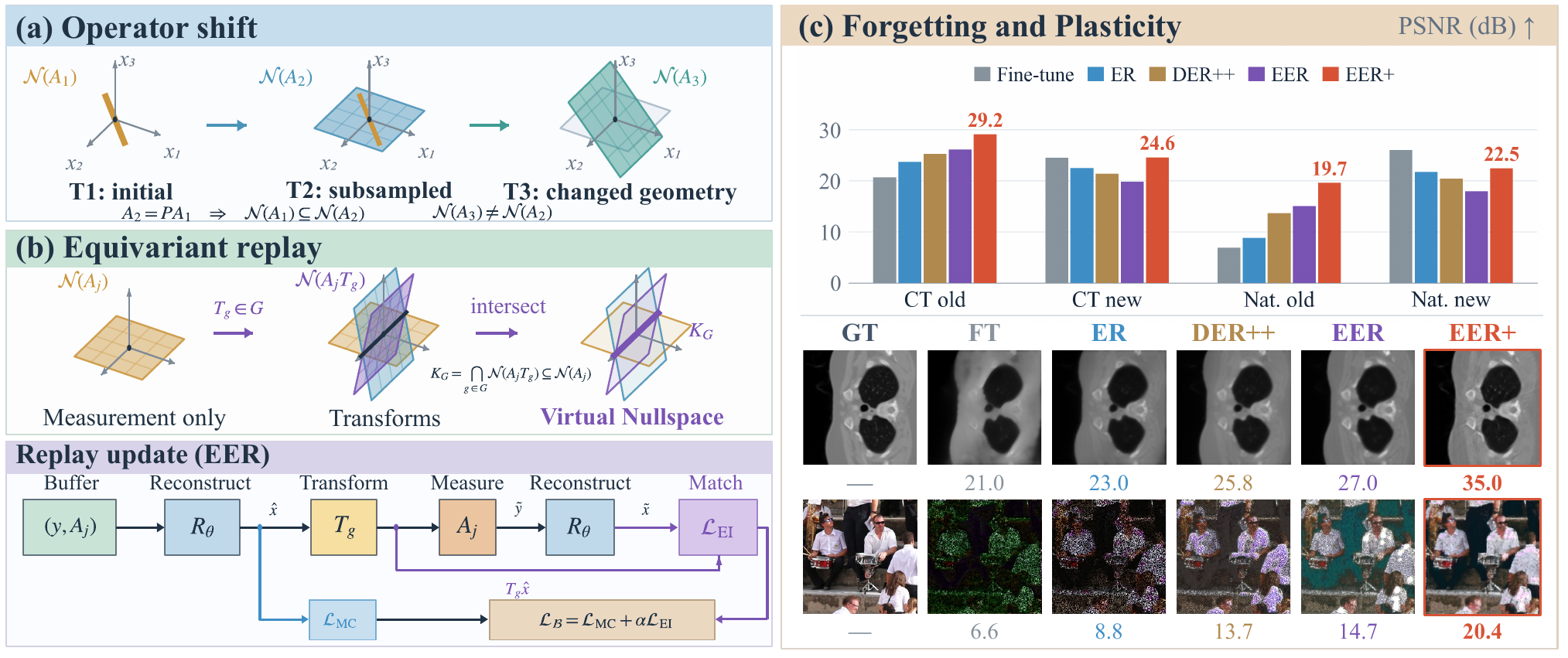}
    
\caption{
\textbf{Equivariant experience replay for continual imaging.}
\textbf{(a) Operator shift.}
Changes in acquisition alter the operator nullspace. The 3D sketches are illustrative.
\textbf{(b) Equivariant replay.}
From a stored pair $(y,A_j)$, EER reconstructs $\hat{x}$, generates $\tilde{y}=A_jT_g\hat{x}$, and reconstructs $\tilde{x}$ with shared parameters. Virtual operators provide complementary constraints;
\textbf{(c) Forgetting and Plasticity.}
Validation PSNR after new task on medical CT imaging and natural image processing. \emph{Old} averages past-task PSNR; \emph{New} denotes current-task PSNR. Qualitative results are past-task reconstructions after learn new task.}

    \label{fig:overview}
\end{figure*}

\paragraph{Challenges and overview.}
The standard self-supervised objective for inverse problems is measurement consistency (MC),
\begin{equation}
    \mathcal{L}_{\mathrm{MC}}(\theta; y, A) = \big\| A\,\mathcal{F}_{\theta}(y, A) - y \big\|_2^2 .
    \label{eq:mc}
\end{equation}
MC constrains only the component of $\hat{x}$ in the range of $A^{\top}$. Since $A$ is typically rank-deficient, any perturbation of $\hat{x}$ within the nullspace $\mathcal{N}(A)$ leaves $\mathcal{L}_{\mathrm{MC}}$ unchanged. Indeed, networks trained with MC alone perform on par with the zero-filled reconstruction~\citep{wang2025fully}.
A natural baseline for our setting combines experience replay (ER)~\citep{rolnick2019experience} with task-dependent MC, 
so that historical measurements are replayed under their respective acquisition and noise models; we refer to this baseline as ER+MC. However, operator shift makes this baseline fragile. Each task has its own nullspace, and fitting the current operators provides no information about how the network should fill in $\mathcal{N}(A)$ for past operators $A \in \mathcal{A}_j$, $j<i$. The nullspace components of past tasks are therefore protected only by the replay term. This term is estimated from at most $K$ measurements, and each measurement constrains only the range component of its image.
To constrain these nullspace components, we propose \emph{physics-guided experience replay}, which uses the stored operators to derive additional self-supervised constraints from each buffered measurement through equivariance. These constraints supervise the nullspace components that MC leaves unconstrained. Fig.~\ref{fig:overview} illustrates the overall framework, and 
Sec.~\ref{subsec:method} details each component.

\subsection{Continual Learning via EER}\label{subsec:method}

\subsubsection{Operator Shift and the Nullspace Problem}\label{sec:method_nullspace}

As discussed in Sec.~\ref{sec:formulation}, MC provides no information about the component of $x$ in $\mathcal{N}(A)$. More generally, self-supervised learning from measurements acquired through a single fixed incomplete operator is impossible, because nothing in the data reveals the nullspace component; the limitation can be lifted only when measurements come from \emph{multiple} operators whose nullspaces differ~\citep{tachella2022unsupervised}.

This observation shows that operator shift plays two roles in continual reconstruction. Across the whole sequence, the operator families $\mathcal{A}_1,\ldots,\mathcal{A}_N$ jointly provide the operator diversity that makes self-supervised learning possible. At any given stage $i$, however, this diversity is no longer available: the operators of past tasks survive only through the at most $K$ pairs $(y,A)$ stored in the buffer. Under operator shift, forgetting is therefore not only a drift of the network parameters; it is also a loss of the operator diversity that kept the nullspace components of past tasks identifiable. 
This insight motivates a principled approach to design an effective replay mechanism that explicitly recovers this diversity from a small buffer, which we address next by extracting multiple constraints from each individually stored measurement.

\subsubsection{Virtual Operators from Invariance}
\label{sec:method_virtual}
Let $\mathcal{G}$ be a group of transformations $\{T_g\}_{g\in\mathcal{G}}$ (e.g., rotations or shifts) under which the image distribution is invariant, i.e., $T_g x \sim p_j$ whenever $x \sim p_j$. Any measurement then admits the factorization
\begin{equation}
    y = A x = (A T_g)\,(T_g^{-1} x), \qquad T_g^{-1}x \sim p_j .
    \label{eq:virtual}
\end{equation}
The same $y$ can thus be read as an observation of an equally plausible image $T_g^{-1}x$ through the \emph{virtual operator} $A T_g$. Because $\mathcal{N}(A T_g) = T_g^{-1}\mathcal{N}(A)$ differs from $\mathcal{N}(A)$ for non-trivial $g$, the orbit $\{A T_g\}_{g\in\mathcal{G}}$ supplies the operator diversity that a single stored pair lacks. Two existing self-supervised losses exploit~\eqref{eq:virtual} in different ways.

\paragraph{Equivariant imaging (EI).}
EI~\citep{chen2021equivariant} applies the virtual operators to the reconstruction itself. Given $\hat{x} = \mathcal{F}_\theta(y,A)$, it generates a virtual measurement $A T_g \hat{x}$ and requires the network to recover $T_g\hat{x}$ from it:
\begin{equation}
    \mathcal{L}_{\mathrm{EI}}(\theta; y, A)
    = \mathbb{E}_{g}\Big[\big\| \mathcal{F}_\theta(A T_g \hat{x}, A) - T_g \hat{x} \big\|_2^2\Big].
    \label{eq:ei}
\end{equation}
While EI is typically employed alongside $\mathcal{L}_{\mathrm{MC}}$, we note a design property that affects its reliability~\citep{xu2025fast}: its supervision signal is generated from the network's own output $\hat{x}$, so its quality depends on that of the current reconstruction.

\paragraph{Equivariant splitting (ES).}
ES~\citep{sechaud2026equivariant} instead uses the virtual operators on the measurement side, and evaluates the network only against acquired data. Let $M_1$ be a random splitting mask on the measurement space and $M_2 = I - M_1$. For $g \in \mathcal{G}$, define
\begin{equation}
    y_k = M_k y, \qquad A_{g,k} = M_k A T_g, \qquad k \in \{1,2\},
\end{equation}
and reconstruct from the first subset, $\hat{x}_{g} = \mathcal{F}_\theta(y_1, A_{g,1})$, which by~\eqref{eq:virtual} is an estimate of $T_g^{-1}x$. For noiseless measurements, the ES loss is
\begin{equation}
    \mathcal{L}_{\mathrm{ES}}(\theta; y, A)
    = \mathbb{E}_{g, M_1}\Big[
        \big\| A_{g,1}\hat{x}_{g} - y_1 \big\|_2^2
        + \big\| A_{g,2}\hat{x}_{g} - y_2 \big\|_2^2
    \Big],
    \label{eq:es}
\end{equation}
where the first term enforces consistency with the input subset and the second predicts the held-out subset. ES additionally requires the reconstruction network to be equivariant in the sense
\begin{equation}
    \mathcal{F}_\theta(y, A T_g) = T_g^{-1}\, \mathcal{F}_\theta(y, A), \qquad \forall g \in \mathcal{G},
    \label{eq:equiv}
\end{equation}
which any backbone satisfies after group averaging, $\mathcal{F}^{\mathcal{G}}_\theta(y,A) = \frac{1}{|\mathcal{G}|}\sum_{g} T_g\, \mathcal{F}_\theta(y, A T_g)$. 
Under~\eqref{eq:equiv}, the splitting loss is an unbiased estimate of the supervised loss~\citep{sechaud2026equivariant}. 

\subsubsection{Equivariant Experience Replay}
\label{sec:method_eer}
We now use the virtual operators of Sec.~\ref{sec:method_virtual} to restore, at replay time, the operator diversity identified in Sec.~\ref{sec:method_nullspace}. Write the buffer as $\mathcal{B}_{i-1} = \bigcup_{j<i}\mathcal{B}_j$, where $\mathcal{B}_j$ holds the pairs $(y,A)$ retained from task $j$, 
and for any per-sample loss $\mathcal{L}_q$ and set $\mathcal{S}$ let $\mathcal{L}_q^{\mathcal{S}} = \frac{1}{|\mathcal{S}|}\sum_{(y,A)\in\mathcal{S}}\mathcal{L}_q(\theta;y,A)$. The replay loss averages over past tasks,
\begin{equation}
    \mathcal{L}_q^{\mathcal{B}_{<i}} = \frac{1}{i-1}\sum_{j=1}^{i-1}\mathcal{L}_q^{\mathcal{B}_j}.
    \label{eq:replay}
\end{equation}

\paragraph{Replay under operator shift.}
Equivariant self-supervision was developed for a \emph{single, fixed} incomplete observation model~\citep{chen2021equivariant,sechaud2026equivariant}. In our setting, the operators may vary across sequential tasks: at stage $i$, the network is trained on a mixture of current operators $A\in\mathcal{A}_i$ and a small set of stored past operators, and the latter can never be resampled. Applying equivariance to the buffer turns each stored operator into its orbit, so that the effective operator set seen during replay is
\begin{equation}
    \Omega_i = \big\{ A T_g \,:\, (y,A) \in \mathcal{B}_{i-1},\ g \in \mathcal{G} \big\},
    \label{eq:orbit}
\end{equation}
which is a factor of up to $|\mathcal{G}|$ larger than the stored set and whose nullspaces jointly cover more of $\mathcal{N}(A)$ for each past $A$. Each buffered pair therefore yields a whole family of constraints, one per virtual operator and, for ES, one per random split.


\paragraph{Training objectives.}
All variants share the buffer and differ only in the self-supervised losses. With the current task-specific self-supervised loss
\begin{equation}
    \mathcal L_{\mathrm{SS},j}=\mathcal L_{\mathrm{obs},j}+\lambda_j\mathcal R_j,
\end{equation}
where $\mathcal L_{\mathrm{obs},j}$ is an observation-based loss adapted to the acquisition and noise model, and $\mathcal R_j$ is an optional self-supervised regularizer. For example, $\mathcal L_{\mathrm{obs},j}$ may be MC for noiseless incomplete measurements or R2R for Gaussian denoising. With the current-task objective
\begin{equation}
\mathcal L_i^{\mathrm{new}}
=
\mathcal L_{\mathrm{SS},i}^{D_i}
+
\eta_i\mathcal L_{\mathrm{EI},i}^{D_i},
\label{eq:current_loss}
\end{equation}
we consider
\begin{align}
    \mathcal L_i^{\mathrm{ER+SS}}&=\mathcal L_i^{\mathrm{new}}+\frac{\beta}{i-1}\sum_{j<i}\mathcal L_{\mathrm{SS},j}^{B_j}, \label{eq:ermc}\\
    \mathcal L_i^{\mathrm{EER}}&=\mathcal L_i^{\mathrm{new}}+\frac{\beta}{i-1}
    \sum_{j<i}
    \left(
    \mathcal L_{\mathrm{SS},j}^{B_j}+
    \eta_j\mathcal L_{\mathrm{EI},j}^{B_j}
    \right), \label{eq:eer}\\
    \mathcal L_i^{\mathrm{EER+}}&=\mathcal L_{\mathrm{ES},i}^{D_i}+\frac{\beta}{i-1}\sum_{j<i}\mathcal L_{\mathrm{ES},j}^{B_j}, \label{eq:eerplus}
\end{align}
where $\beta$ balances forgetting against plasticity. ER+SS replays historical measurements using their respective self-supervised objectives, with ER+MC as its noiseless MC-based instantiation. EER adds EI to the replay term, so that buffered measurements are also seen through virtual operators, and differs from ER+MC only in the replay loss. EER+ uses ES on both current and buffered data. 

\subsection{Why Equivariant Replay Retains Past Tasks}
\label{sec:eer_risk_calibration}

At stage $i$, the reconstruction error on a past task $j$ cannot be computed, because its ground truth is unavailable; the learner only observes the replay loss on $\mathcal{B}_j$. Replay can therefore prevent forgetting only if \emph{a small replay loss implies a small past-task error}. MC replay does not satisfy this condition: adding any $u\in\mathcal{N}(A_j)$ to a reconstruction leaves $\mathcal{L}_{\mathrm{MC}}$ unchanged (Sec.~\ref{sec:method_nullspace}), so the replay loss can remain minimal while the error on task $j$ grows. We show that the ES replay loss of EER+ satisfies this condition, and that operator shift determines how tightly it does so. Proofs of Proposition~\ref{prop:main_es_calibration} are given in supplementary material. 


\paragraph{Replay risk and coverage.}
Fix a past task $j$ with noiseless measurements $y=A_jx$, where $x\sim p_j$. For a random split $M$ and transformation $T_g$, let $v=(My,MA_jT_g)$. The population ES replay risk is
\begin{equation}
    \mathcal{R}_j(\mathcal{F})
    =\mathbb{E}_{x,g,M}
    \left\|
    W_M^{1/2}
    \left(A_jT_g\mathcal{F}(v)-y\right)
    \right\|_2^2,
\end{equation}
where $W_M$ weights the input and held-out residuals as in the normalized ES loss. Following the averaged coverage formulation, define
\begin{equation}
\begin{aligned}
    Q_j(S)
    &=\mathbb{E}\!\left[
    (A_jT_g)^\top W_M(A_jT_g)
    \mid MA_jT_g=S
    \right],\\
    \bar Q_j&=\mathbb{E}_M[Q_j(MA_j)],
    \qquad
    \bar\mu_j=\lambda_{\min}(\bar Q_j).
\end{aligned}
\end{equation}
For $\bar\mu_j>0$, define the coverage-weighted aggregate
\begin{equation}
    \bar{\mathcal{F}}_j(y)
    =\bar Q_j^{-1}
    \mathbb{E}_M
    \left[
    Q_j(MA_j)\mathcal{F}(My,MA_j)
    \right].
\end{equation}
Let $\bar{\mathcal{F}}_j^\star$ denote the corresponding aggregate of the MMSE reconstructor
$\mathcal{F}_j^\star(z,S)=\mathbb{E}[x\mid Sx=z]$,
and define
$D_j(\mathcal{F})=
\mathbb{E}_y\|\bar{\mathcal{F}}_j(y)-
\bar{\mathcal{F}}_j^\star(y)\|_2^2$.

\begin{proposition}[Replay loss controls past-task error]
\label{prop:main_es_calibration}
Assume that (A1) $p_j$ is invariant under the unitary transformations $\{T_g\}_{g\in\mathcal{G}}$, and (A2) $\bar\mu_j>0$. Then, for every equivariant reconstructor $\mathcal{F}$,
\begin{equation}
    D_j(\mathcal{F})
    \leq
    \frac{\mathcal{R}_j(\mathcal{F})-\mathcal{R}_j^\star}
    {\bar\mu_j}.
    \label{eq:replay_calibration}
\end{equation}
\end{proposition}

Setting $\mathcal R_j(F)=\mathcal R_j^\star$ recovers the ES
result: $\bar F_j=\bar F_j^\star$ almost surely. We further develop the upper bound beyond the optimum, which is the regime of continual learning: the past-task error is controlled by the replay loss, with stronger operator coverage yielding a tighter bound. For MC replay, $\mu_j=0$ and this guarantee is lost.


\section{Experiments}\label{sec:exp}
We isolate the effect of operator shift by keeping the image distribution fixed throughout continual learning and varying only the forward operator. Within this setting, we answer three questions.
\textbf{(Q1)} \emph{Does operator shift alone cause forgetting under self-supervision, and how does it depend on the relation between the previous and current operators?} We contrast nested and non-nested shifts in CT, as well as a cross-family shift in imaging processing, where the new operator carries no information about the old nullspace.
\textbf{(Q2)} \emph{With a single stored measurement, does expanding its operator into an orbit of virtual operators preserve past-task reconstruction better than replaying the operator alone (ER+MC) or distilling past outputs (DER, DER++)?} 
\textbf{(Q3)} \emph{Can equivariant replay retain knowledge across multiple operator shifts while adapting to a new operator?} We measure forgetting and new-task performance across multiple-task sequences, together with training and inference costs.

\subsection{Benchmarks and protocol}
\label{sec:exp_setup}

\paragraph{Two regimes of operator shift.}
Both benchmarks use disjoint training images from the same underlying distribution and consider two regimes:
\begin{itemize}
    \item \emph{Within-family operator shifts} (CT). The first shift reduces projection angles while preserving a subset of the previous measurements, yielding $\mathcal{N}(A_1)\subseteq\mathcal{N}(A_2)$. The second introduces limited-angle acquisition with a different nullspace structure. The network must therefore adapt to successive changes in the operator without degrading its reconstructions on previous tasks.

    \emph{Cross-family shift} (natural images). The operator changes from pixel inpainting to Gaussian denoising. The denoising operator $A_{\mathrm{new}} = I$ has a trivial nullspace, so training on $\mathcal{T}_3$ provides no signal at all about how to fill $\mathcal{N}(A_{\mathrm{past}})$. Inpainting performance can then be retained only through replay, which makes this benchmark the most direct test of the argument in Sec.~\ref{sec:method_nullspace}.
\end{itemize}

\paragraph{Sparse-view CT.}
We use LoDoPaB-CT~\citep{leuschner2021lodopab}, resizing slices from $362^2$ to $128^2$. Each task uses 256 training slices from 128 patients, with 64 slices from 32 additional patients for validation. All tasks use noiseless parallel-beam tomography with 182 detector pixels. $\mathcal{T}_1$ uses 100 uniformly spaced views over $[0^\circ,180^\circ)$, $\mathcal{T}_2$ retains 20 of these views, and $\mathcal{T}_3$ uses 60 views over $[0^\circ,120^\circ)$.

\paragraph{Natural images.}
We use DIV2K~\citep{agustsson2017ntire}, with one $256\times256$ RGB crop per image, 100 training images per task, and 50 validation images. $\mathcal{T}_1$ and $\mathcal{T}_2$ use noiseless inpainting with observed-pixel fractions of 0.4 and 0.2, respectively; $\mathcal{T}_1$ uses a fixed Bernoulli mask shared across color channels. $\mathcal{T}_3$ performs Gaussian denoising with $\sigma=25/255$, using one fixed noise realization per training image.

\paragraph{Continual protocol.}
The model is first trained on $\mathcal{T}_1$ and then on $\mathcal{T}_2$, followed by $\mathcal{T}_3$ for both benchmarks, without access to previous training sets. Replay-based methods keep one randomly selected measurement from each previous task together with its forward operator, giving a buffer of $K=1$ for two tasks and $K=2$ at the third stage. The operator is known at test time. All continual methods train without ground truth, using a shared externally supervised initialization~\citep{chen2021equivariant}. Ground truth is used only for validation and the supervised Joint reference.

\paragraph{Metrics.}
Let $a_{i,j}$ denote the validation PSNR on task $\mathcal{T}_j$ after learning $\mathcal{T}_i$. For two-task experiments, we report new-task PSNR $a_{2,2}$, old-task PSNR $a_{2,1}$, and define forgetting as $\mathrm{Fgt}=a_{1,1}-a_{2,1}$.
For three-task sequences, we additionally report average PSNR, backward transfer (BWT), and learning accuracy (LA):
\begin{equation}
    \mathrm{BWT}=\frac{1}{T-1}\sum_{j=1}^{T-1}
    (a_{T,j}-a_{j,j}),\qquad
    \mathrm{LA}=\frac{1}{T}\sum_{j=1}^{T}a_{j,j}.
    \label{eq:cl_metrics}
\end{equation}
We also report per-task forgetting across stages. Each method uses a fixed inference procedure, including at the shared $\mathcal{T}_1$ parameters. EER+ is evaluated with $S=1$ and $S=10$ splits to distinguish single-split retention from averaging gains. Computational cost is measured by $\mathcal{T}_3$ training time and per-image inference latency.

\subsection{Compared Methods}
\label{sec:exp_methods}
All continual methods start from the same $\mathcal{T}_1$ model, share the same measurement buffer, and optimize
\begin{equation}
    \mathcal{L} = \mathcal{L}_{\mathrm{new}} + \beta\, \mathcal{L}_{\mathrm{rep}}, \qquad \beta = 4,
\end{equation}
where $\mathcal{L}_{\mathrm{new}}$ is MC+EI~(\eqref{eq:ei}) for every method except EER+. \textbf{Existing CL methods assume labeled replay -- we adapt them to the measurement-only setting by replacing every supervised term with its closest self-supervised counterpart:}

\paragraph{Baselines and proposed methods.}
\textbf{Fine-tune} \citep{darestani2022test} trains on current task without replay ($\beta = 0$) and indicates the amount of forgetting that occurs without any protection.
\textbf{ER+MC}~\citep{rolnick2019experience} is the baseline of Sec.~\ref{sec:formulation}. Since no labels are available, it replays the buffer with measurement consistency, $\mathcal{L}_{\mathrm{rep}} = \mathcal{L}_{\mathrm{MC}}^{\mathcal{B}}$.
\textbf{DER}~\citep{buzzega2020dark} distills the outputs of the $\mathcal{T}_1$ model. We replace logit matching with image matching. At the task boundary, we store $z_b = \mathcal{F}_{\theta_1}(y_b, A_b)$ for each buffered pair $(y_b, A_b)$ and minimize
\begin{equation}
    \mathcal{L}_{\mathrm{DER}}^{\mathcal{B}} = \mathbb{E}_{(y_b, A_b, z_b) \sim \mathcal{B}} \big\| \mathcal{F}_\theta(y_b, A_b) - z_b \big\|_2^2 .
\end{equation}
DER thus stores one reconstruction in addition to each buffered measurement.
\textbf{DER++}~\citep{buzzega2020dark} combines distillation with replay. We replace the supervised replay term of the original method with measurement consistency, $\mathcal{L}_{\mathrm{rep}} = \mathcal{L}_{\mathrm{MC}}^{\mathcal{B}} + \gamma\, \mathcal{L}_{\mathrm{DER}}^{\mathcal{B}}$, with $\gamma = 1$.
\textbf{EER \& EER+} are our methods (Sec.~\ref{sec:method_eer}). EER replays MC+EI~\eqref{eq:eer}. EER+ uses equivariant splitting for both the current and the replay term~\eqref{eq:eerplus}, and at inference it averages $S$ split reconstructions (Sec.~\ref{sec:exp_impl}).
\textbf{Joint (sup.)} trains on the complete training sets of both tasks with an equally weighted supervised MSE loss. It uses clean images and all the data.

\paragraph{Memory.}
All replay methods store the same measurement-operator pair. In both benchmarks, the $\mathcal{T}_1$ operator is a fixed geometry. DER and DER++ additionally store one image per buffered measurement, so they use more memory than ER+MC, EER, and EER+.

\paragraph{Self-supervised loss for denoising.}
For $\mathcal{T}_3$ on natural images, $A = I$ and MC reduces to $\| \mathcal{F}_\theta(y, I) - y \|_2^2$. This loss is minimized by the identity map and therefore provides no denoising signal. For every method, we replace MC on the current task with Recorrupted-to-Recorrupted (R2R)~\citep{pang2021recorrupted},
\begin{equation}
    \mathcal{L}_{\mathrm{R2R}} = \big\| \mathcal{F}_\theta(y + \omega, I) - (y - \omega) \big\|_2^2, \qquad \omega \sim \mathcal{N}(0, \sigma^2 I),
\end{equation}
which corresponds to the recorruption strength $\alpha = 0.5$ of~\cite{pang2021recorrupted}. All methods except EER+ use R2R+EI as $\mathcal{L}_{\mathrm{new}}$. EER+ uses the split variant of R2R (Sec.~\ref{sec:exp_impl}). Replay on the noiseless inpainting buffer is unchanged.

\subsection{Implementation Details}
\label{sec:exp_impl}

\paragraph{Architecture and initialization.}
All methods use the residual U-Net of~\citep{chen2021equivariant}, with $\mathcal{F}_\theta(y,A)=f_\theta(A^\dagger y)$, where $A^\dagger$ denotes filtered back-projection for CT, zero-filling for inpainting, and the identity for denoising. We initialize from the pretrained 50-view CT or Urban100 inpainting (mask rate 0.3) models of~\citep{chen2021equivariant}. Training on $\mathcal{T}_1$ uses MC+EI for 40 epochs with Adam, learning rate $10^{-4}$, weight decay $10^{-8}$, and batch size 2. At $\mathcal{T}_2$, a fresh Adam optimizer uses learning rate $5\times10^{-4}$ with the same weight decay and batch size. Training runs for 40 epochs unless stated otherwise, with buffered measurements repeated to fill each batch. EER+ uses 20 epochs in CT, while Joint is trained for 80 epochs on each benchmark.


\paragraph{Transformations and equivariant splitting.}
For natural images, EI and ES use circular shifts, sampling one shift per iteration. For CT, EI uses four random rotations with bilinear interpolation, while ES uses the dihedral group of right-angle rotations and reflections. Equivariant splitting retains each measurement entry with probability 0.6, applying the mask $M$ to projection angles in CT or observed pixels in natural images, shared across color channels. The ES loss normalizes the input and held-out residuals by their respective measurement fractions. The masked back-projection has a closed form for pixel masks; for CT, we approximate it using 20 conjugate-gradient iterations.

\paragraph{Inference and implementation.}
For denoising, split R2R applies the mask to $y+\omega$ and evaluates residuals against $y-\omega$. At inference, EER+ averages $S\in\{1,10\}$ split reconstructions without recorruption, denoted S-1 and S-10; all other methods use single-pass inference. Experiments use deepinv 0.3.5~\cite{tachella2025deepinverse} on an NVIDIA RTX 4060 Ti GPU.

\subsection{Main Results}
\label{sec:exp_main}

\paragraph{Operator shift alone causes severe forgetting (Q1).}
Although the image distribution is unchanged, Fine-tune loses averaged 9.95\,dB on the past tasks in CT and 17.58\,dB in natural images as shown in Table~\ref{tab:three_task_forgetting} and~\ref{tab:three_task_natural_forgetting}. Under the cross-family shift, the denoising operator has a trivial nullspace, so $\mathcal{T}_3$ provides no signal about how to fill $\mathcal{N}(A_{\mathrm{past}})$, and inpainting performance collapses to 7.24\,dB on $\mathcal{T}_1$ and 6.60\,dB on $\mathcal{T}_2$, as presented in Table~\ref{tab:three_task_natural}. The CT shift is less destructive but still substantial. Forgetting in continual inverse problems is therefore not confined to shifts in the image distribution; a change of operator suffices.

\paragraph{Replaying the stored operator alone barely helps.}
ER+MC reduces forgetting by only 2.30\,dB in CT and 2.60\,dB in natural images. This is the failure predicted in Sec.~\ref{sec:formulation}: MC on the stored inpainting measurement is satisfied by any reconstruction that agrees with it on the observed pixels, and therefore places no constraint on $\mathcal{N}(A_{\mathrm{past}})$. Distillation helps only marginally, although DER and DER++ store an additional image per measurement (BWT -6.07/-6.18\,dB in CT, -11.89/-10.40\,dB in natural images). Matching a single stored output constrains the network at a single input and does not generalize to the rest of the old task.

\paragraph{Expanding the stored operator into an orbit preserves the past tasks (Q2).}
Both equivariant variants retain more than every baseline on both benchmarks. EER lowers BWT relative to ER+MC by 2.43\,dB in CT and 5.77\,dB in natural images. EER+ goes much further: with single-pass inference (S-1), which measures retention alone, its BWT is 0.22\,dB in CT and 1.75\,dB in natural images. This removes 98\% and 90\% of the forgetting of Fine-tune, and is 5.00 and 7.46\,dB lower than EER on both benchmarks. Averaging ten splits (S-10) raises PSNR of sequential tasks further while at a cost of computation.

\paragraph{Retention versus plasticity (Q3).}
EER+ preserves past tasks while maintaining strong adaptation to new operators. In three-task CT, EER+ (S-1) achieves 24.62\,dB on the new task, outperforming all competing continual methods while improving BWT from $-9.95$ to $-0.22$\,dB (Tables~\ref{tab:three_task_forgetting} and~\ref{tab:three_task_ct}). On natural images, it achieves 22.48\,dB on denoising, exceeding all replay baselines while reducing forgetting to 1.75\,dB (Tables~\ref{tab:three_task_natural_forgetting} and~\ref{tab:three_task_natural}). Averaging ten splits further raises the average PSNR to 29.27\,dB in CT and 21.62\,dB in natural images. These results show that equivariant replay enables effective adaptation to successive operator shifts without sacrificing retention.

\paragraph{Computational cost.}
EER+ requires less training time than ER+MC and EER on both benchmarks (Tables~\ref{tab:three_task_forgetting} and~\ref{tab:three_task_natural_forgetting}), while storing only one measurement-operator pair per past task. Its main overhead lies in inference: S-1 takes 51.1\,ms per image in CT and 9.7\,ms in natural images, while S-10 increases latency in exchange for higher reconstruction accuracy. EER+ thus offers a tradeoff between inference cost and reconstruction quality.

\begin{table}[t]
\centering
\caption{
Forgetting, continual learning (CL) metrics, and training cost in the three-task CT experiment. Training time is the wall-clock time of the $\mathcal{T}_3$ stage.
EER+ S-1 and S-10 share the same model. The two CT100 columns measure forgetting of $\mathcal{T}_1$ from different reference
points: since $\mathcal{T}_1$ ($a_{1,1}-a_{3,1}$, cumulative over $\mathcal{T}_2$ and $\mathcal{T}_3$) and
since $\mathcal{T}_2$ ($a_{2,1}-a_{3,1}$, caused by $\mathcal{T}_3$ only).
}
\label{tab:three_task_forgetting}
\scriptsize
\setlength{\tabcolsep}{2.5pt}
\resizebox{\linewidth}{!}{
\begin{tabular}{llrrrrrrr}
\toprule
& &
\multicolumn{3}{c}{Forgetting (dB) $\downarrow$}
& \multicolumn{2}{c}{CL metrics (dB)}
& \multicolumn{2}{c}{Cost} \\
\cmidrule(lr){3-5}
\cmidrule(lr){6-7}
\cmidrule(lr){8-9}
Method & Inf.
& \shortstack{CT100\\$\mathcal{T}_1\!\to\!T_3$}
& \shortstack{CT100\\$\mathcal{T}_2\!\to\!T_3$}
& \shortstack{CT20\\$\mathcal{T}_2\!\to\!T_3$}
& BWT $\uparrow$ & LA $\uparrow$
& \shortstack{Train\\(s)} & \shortstack{Infer\\(ms/img)} \\
\midrule
Fine-tune & F
& 14.42 & 7.26 & 5.47
& $-9.95$ & 28.62
& 596.1 & 3.3 \\
ER+MC & F
& 11.99 & 6.58 & 3.32
& $-7.65$ & 28.42
& 852.0 & 3.3 \\
DER & F
& 9.22 & 4.56 & 2.92
& $-6.07$ & 27.95
& 725.3 & 3.3 \\
DER++ & F
& 9.43 & 4.91 & 2.93
& $-6.18$ & 28.11
& 715.1 & 3.2 \\
EER (ours) & F
& 8.60 & 4.20 & 1.84
& $-5.22$ & 27.56
& 1516.9 & 3.3 \\
EER+ (ours) & S-1
& $-0.41$ & 0.02 & \textbf{0.84}
& $-0.22$ & 27.79
& 890.3 & 51.1 \\
EER+ (ours) & S-10
& \textbf{$-$1.07} & \textbf{$-$0.42} & 1.03
& $\mathbf{0.02}$ & \textbf{29.27}
& 890.3 & 536.7 \\
\midrule
Joint (sup.) & F
& -- & -- & --
& -- & --
& 2213.9 & 3.6 \\
\bottomrule
\end{tabular}}
\end{table}

\begin{table}[t]
\centering
\caption{
Three-task continual CT reconstruction
(100-view full-angle $\rightarrow$ 20-view full-angle
$\rightarrow$ 60-view limited-angle), evaluated after learning $\mathcal{T}_3$.
CT100, CT20, and LA60 denote $\mathcal{T}_1$, $\mathcal{T}_2$, and $\mathcal{T}_3$.
Avg is the mean PSNR over the three tasks.
F denotes standard single-pass inference; 
S-1 and S-10 denote EER+ with one and ten split reconstructions at inference.
Best continual results are shown in bold.
}
\label{tab:three_task_ct}
\scriptsize
\setlength{\tabcolsep}{3pt}
\resizebox{\linewidth}{!}{
\begin{tabular}{llccccccc}
\toprule
& &
\multicolumn{4}{c}{PSNR (dB) $\uparrow$}
& \multicolumn{3}{c}{SSIM $\uparrow$} \\
\cmidrule(lr){3-6}
\cmidrule(lr){7-9}
Method & Inf.
& CT100 & CT20 & LA60 & Avg
& CT100 & CT20 & LA60 \\
\midrule
Fine-tune & F
& 21.93 & 19.52 & 24.53 & 21.99
& 0.6761 & 0.5393 & 0.7140 \\
ER+MC & F
& 24.36 & 23.06 & 22.54 & 23.32
& 0.8236 & 0.5770 & 0.6758 \\
DER & F
& 27.13 & 23.78 & 20.79 & 23.90
& 0.8391 & 0.6060 & 0.6276 \\
DER++ & F
& 26.92 & 23.67 & 21.39 & 23.99
& 0.8467 & 0.6058 & 0.6490 \\
EER (ours) & F
& 27.76 & 24.65 & 19.84 & 24.08
& 0.8655 & 0.6265 & 0.5994 \\
EER+ (ours) & S-1
& 33.99 & 24.31 & 24.62 & 27.64
& 0.9024 & 0.6189 & 0.7098 \\
EER+ (ours) & S-10
& \textbf{36.95} & \textbf{25.97}
& \textbf{24.94} & \textbf{29.28}
& \textbf{0.9514} & \textbf{0.7173}
& \textbf{0.7468} \\
\midrule
Joint (sup.) & F
& 42.52 & 30.51 & 28.61 & 33.88
& 0.9815 & 0.8287 & 0.8677 \\
\bottomrule
\end{tabular}}
\end{table}

\begin{figure}[t]
    \centering
    \includegraphics[width=0.95\linewidth]{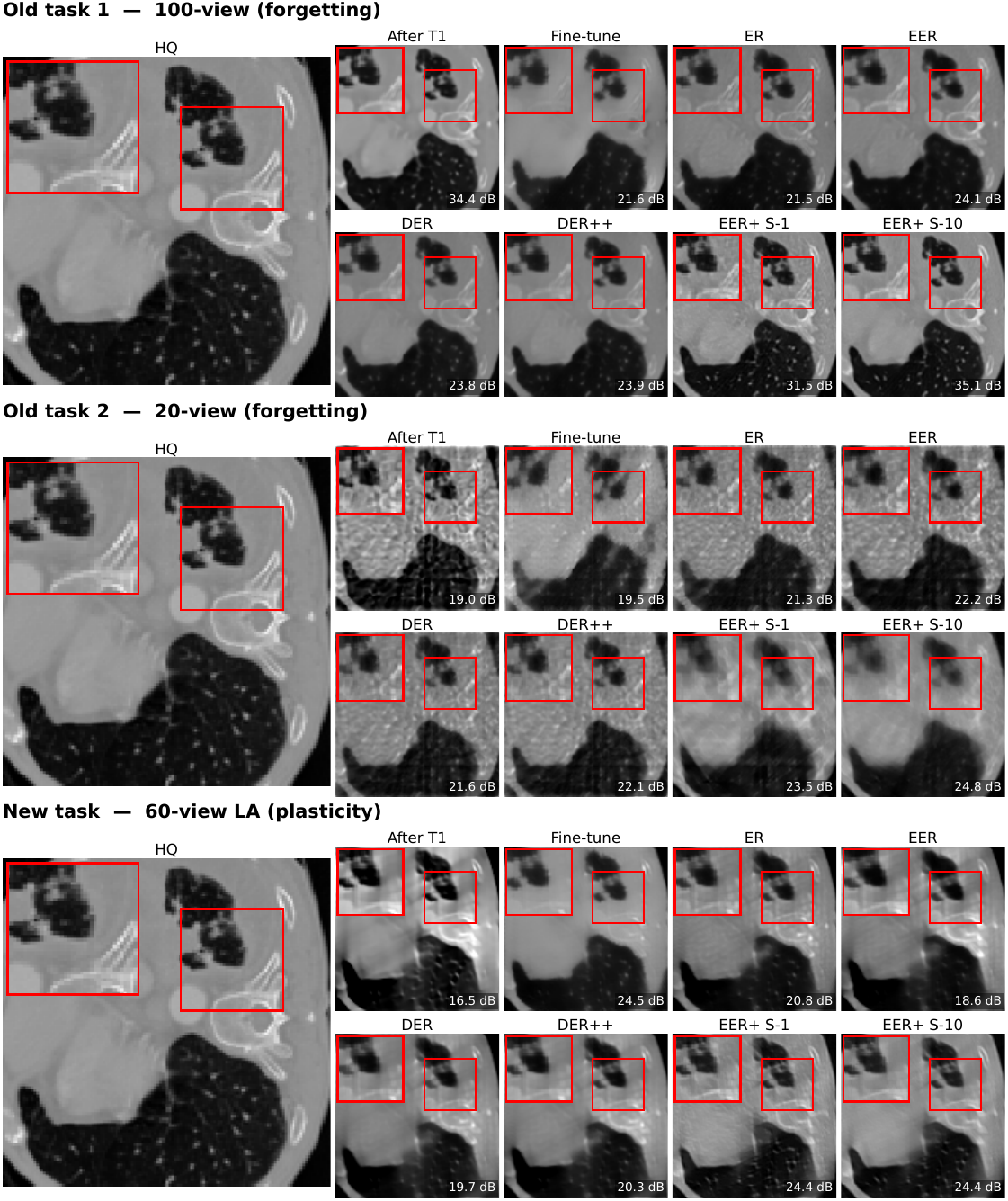}
    \caption{\textbf{Reconstructions after training on $\mathcal{T}_3$ in sparse-view CT.} Top and middle: past tasks (100 and 20 views), showing forgetting. Bottom: new task (60 limited angles), showing plasticity. \emph{After $\mathcal{T}_1$} is the shared $\mathcal{T}_1$ model; the remaining columns are continual methods after $\mathcal{T}_3$. EER+ is shown with S-1 and S-10; all other methods use single-pass inference.}
    \label{fig:main_ct}
\end{figure}

\begin{table}[t]
\centering
\caption{
Forgetting, continual learning (CL) metrics, and cost in the three-task
natural-image experiment.
}
\label{tab:three_task_natural_forgetting}
\scriptsize
\setlength{\tabcolsep}{2.5pt}
\resizebox{\linewidth}{!}{
\begin{tabular}{llrrrrrrr}
\toprule
& &
\multicolumn{3}{c}{Forgetting (dB) $\downarrow$}
& \multicolumn{2}{c}{CL metrics (dB)}
& \multicolumn{2}{c}{Cost} \\
\cmidrule(lr){3-5}
\cmidrule(lr){6-7}
\cmidrule(lr){8-9}
Method & Inf.
& \shortstack{Inp40\\$\mathcal{T}_1\!\to\!T_3$}
& \shortstack{Inp40\\$\mathcal{T}_2\!\to\!T_3$}
& \shortstack{Inp20\\$\mathcal{T}_2\!\to\!T_3$}
& BWT $\uparrow$ & LA $\uparrow$
& \shortstack{Train\\(s)} & \shortstack{Infer\\(ms/img)} \\
\midrule
Fine-tune & F
& 19.24 & 12.93 & 15.93
& $-17.58$ & \textbf{25.02}
& 249.2 & 8.9 \\
ER+MC & F
& 16.74 & 12.43 & 13.21
& $-14.98$ & 23.11
& 487.2 & 8.8 \\
DER & F
& 13.03 & 10.58 & 10.75
& $-11.89$ & 23.49
& 489.6 & 8.9 \\
DER++ & F
& 12.10 & 10.10 & 8.69
& $-10.40$ & 22.87
& 490.9 & 8.9 \\
EER (ours) & F
& 10.30 & 7.09 & 8.12
& $-9.21$ & 22.19
& 735.3 & 8.8 \\
EER+ (ours) & S-1
& \textbf{$-$0.58} & \textbf{1.35} & \textbf{4.08}
& $\mathbf{-1.75}$ & 21.79
& 371.4 & 9.7 \\
EER+ (ours) & S-10
& $-0.18$ & 1.54 & 4.24
& $-2.03$ & 22.98
& 371.4 & 94.0 \\
\midrule
Joint (sup.) & F
& -- & -- & --
& -- & --
& 393.0 & 8.8 \\
\bottomrule
\end{tabular}}
\end{table}

\begin{table}[t]
\centering
\caption{
Three-task continual natural-image restoration
(inpainting with 40\% pixels $\rightarrow$ inpainting with 20\% $\rightarrow$ Gaussian denoising, $\sigma=25/255$),
evaluated after learning $\mathcal{T}_3$.
Inp40, Inp20, and Den denote $\mathcal{T}_1$, $\mathcal{T}_2$, and $\mathcal{T}_3$.
Avg is the mean PSNR over the three tasks.
F denotes standard single-pass inference;
S-1 and S-10 denote EER+ with one and ten split reconstructions at inference.
Best results are shown in bold.
}
\label{tab:three_task_natural}
\scriptsize
\setlength{\tabcolsep}{3pt}
\begin{tabular}{llccccccc}
\toprule
& &
\multicolumn{4}{c}{PSNR (dB) $\uparrow$}
& \multicolumn{3}{c}{SSIM $\uparrow$} \\
\cmidrule(lr){3-6}
\cmidrule(lr){7-9}
Method & Inf.
& Inp40 & Inp20 & Den & Avg
& Inp40 & Inp20 & Den \\
\midrule
Fine-tune & F
& 7.24 & 6.60 & \textbf{26.05} & 13.30
& 0.0787 & 0.0525 & 0.6343 \\
ER+MC & F
& 9.74 & 7.90 & 21.73 & 13.12
& 0.1452 & 0.0874 & 0.4416 \\
DER & F
& 13.45 & 11.36 & 21.87 & 15.56
& 0.2741 & 0.2264 & 0.5456 \\
DER++ & F
& 14.38 & 12.98 & 20.45 & 15.94
& 0.2993 & 0.2594 & 0.5170 \\
EER (ours) & F
& 16.18 & 14.00 & 17.97 & 16.05
& 0.3628 & 0.2945 & 0.4834 \\
EER+ (ours) & S-1
& 22.10 & 17.28 & 22.48 & 20.62
& 0.5988 & 0.4265 & 0.5315 \\
EER+ (ours) & S-10
& \textbf{23.01} & \textbf{17.87} & 23.99 & \textbf{21.62}
& \textbf{0.6714} & \textbf{0.4918} & \textbf{0.6491} \\
\midrule
Joint (sup.) & F
& 28.56 & 25.39 & 28.00 & 27.32
& 0.8353 & 0.7370 & 0.7472 \\
\bottomrule
\end{tabular}
\end{table}

\begin{figure}[t]
    \centering
    \includegraphics[width=0.95\linewidth]{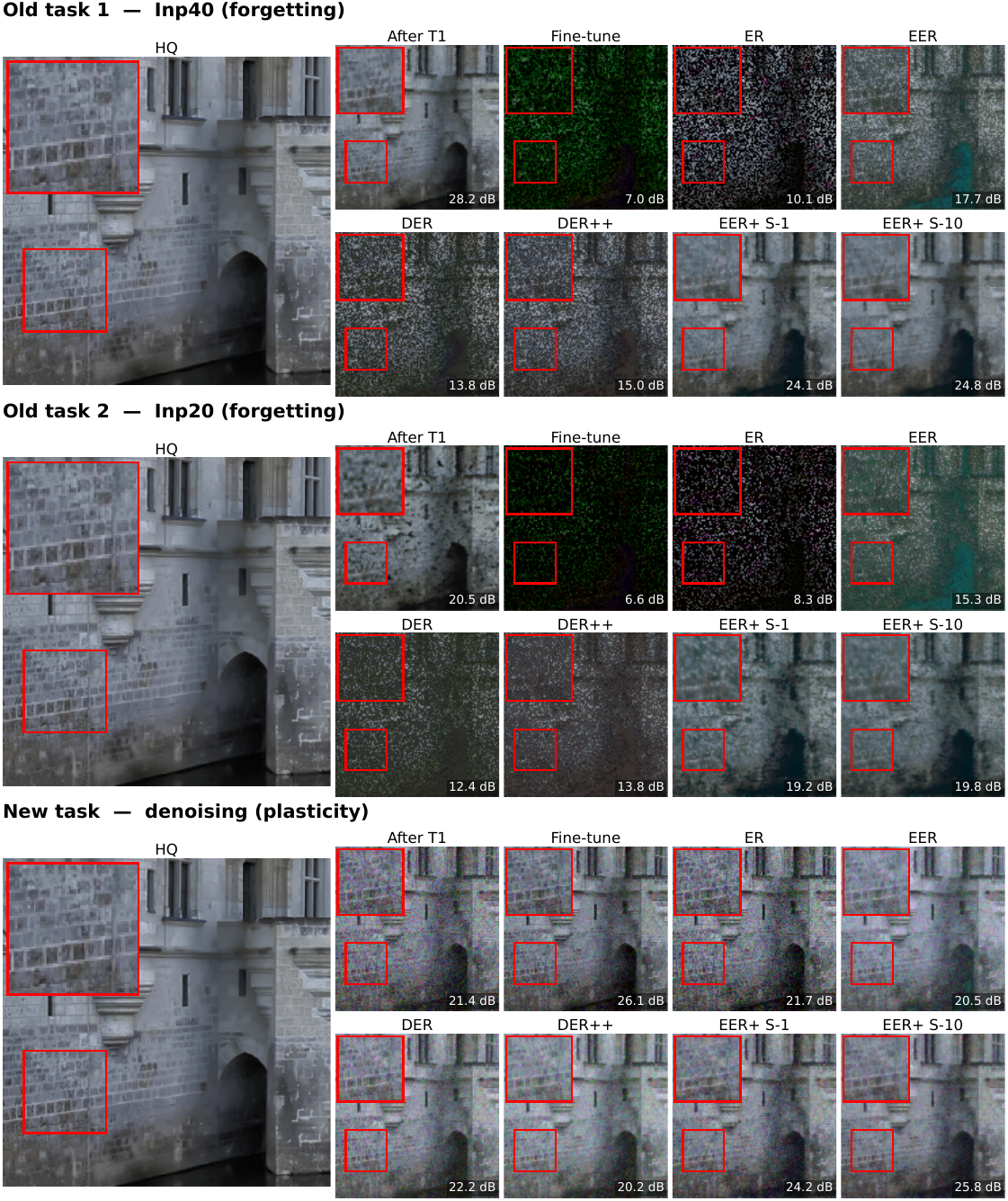}
    \caption{\textbf{Reconstructions after training on $\mathcal{T}_3$ in natural images.} Top and middle: past tasks (inpainting with 40\% and 20\% pixels), showing forgetting. Bottom: new task (denoising), showing plasticity. Layout as in Fig.~\ref{fig:main_ct}.}
    \label{fig:main_nat}
\end{figure}

\section{Ablation Study}
\label{app:replay_ablation}

We isolate the effect of replay supervision by comparing \emph{ES-only} (no replay), \emph{ES+MC} (MC replay), and \emph{EER+} (ours). All variants share the initial model, current-task objective, buffer capacity $K=1$, and $S=10$ inference. We evaluate two operator shifts: 100$\to$20 views on CT and inpainting$\to$denoising on natural images.

\begin{table}[t]
\centering
\caption{
Replay objective ablation. All variants share the current-task loss and $S=10$ inference. Fgt measures the PSNR drop from the initial model (36.35 dB for CT; 26.48 dB for natural images). All metrics are in dB; best results are bold.
}
\label{tab:replay_ablation}
\small
\begin{tabular}{llcccc}
\toprule
Method & Replay
& New $\uparrow$
& Old $\uparrow$
& Fgt $\downarrow$
& Avg $\uparrow$ \\
\midrule
\multicolumn{6}{l}{
\textit{Sparse-view CT: 100 $\rightarrow$ 20 views}
} \\
ES-only & None
& \textbf{27.54} & 32.43 & 3.92 & 29.99 \\
ES+MC & MC
& 27.08 & 35.87 & 0.48 & 31.47 \\
EER+ (ours) & ES
& 26.99 & \textbf{36.53} & \textbf{$-$0.18} & \textbf{31.76} \\
\midrule
\multicolumn{6}{l}{
\textit{Natural images: inpainting $\rightarrow$ denoising}
} \\
ES-only & None
& \textbf{26.51} & 15.62 & 10.86 & 21.07 \\
ES+MC & MC
& 23.69 & 12.49 & 13.99 & 18.09 \\
EER+ (ours) & ES
& 25.58 & \textbf{20.83} & \textbf{5.65} & \textbf{23.21} \\
\bottomrule
\end{tabular}
\end{table}

EER+ consistently depresses forgetting over ES+MC (Table~\ref{tab:replay_ablation}). In CT, it reduces forgetting from 0.48 to $-0.18$\,dB while maintaining comparable new-task accuracy. On natural images, EER+ improves old-task PSNR by 8.34\,dB and new-task PSNR by 1.89\,dB. These results establish the advantage of equivariant replay under limited memory. Three-task extension in supplementary material further confirms stronger retention across operator shifts.
\section{Conclusion}

In this work, we propose Equivariant Experience Replay -- a continual learning framework for imaging inverse problems. Our EER and EER+ schemes are able to significantly mitigate the catastrophic forgetting phenomenon in training deep inverse models pretrained on a different task. To the best of our knowledge, this is the first experience replay method designed specifically for inverse problems utilizing the measurement structure. We therefore also adapt existing experience replay methods into our ground-truth-free CL setting for comparison. We have demonstrated the superiority of our approach over standard methods numerically in medical imaging and image processing applications. It is worth noting that, in principle, our framework is able to joint-force with other regularization and distillation techniques easily, which the practitioners can choose according to their needs. We also observe that it is possible to accelerate our training framework using operator sketching techniques \citep{perelli2021regularization,Oscanoa2024Coil,tang2025practical,xu2026equivariant}, which we leave here as future work.


\section*{LLM Usage Statement}
In this work, we used generative AI tools to check equations and proofs of the results, and suggest corrections when appropriate. The theoretical results of this paper, Proposition~\ref{prop:main_es_calibration}, was proved by the authors, then reviewed and improved with AI assistance (via ChatGPT). The tools suggested improvements to the arguments, some of which we adopted. Every adopted improvement was verified by the authors before being incorporated, and none of them changed the statements of the results. AI was also used to write and review the code of our experiments (via ChatGPT and Grok) as follows: implementing model structures, assisting hyperparameters-tuning, writing scripts to launch training runs and to save and retrieve checkpoints and results, generating figures from experimental data. Additionally, we used generative AI tools to proofread and improve the writing of the manuscript (including correcting grammar mistakes, suggesting rephrasing to improve clarity and fluency, compressing the content, drafting some paragraphs from bullet points, via Claude Code and ChatGPT), to identify relevant literature, and to source or search for information. Figure~\ref{fig:overview}(a), (b), which are illustrative, were created using AI from a hand drawing by the authors (via ChatGPT). We have reviewed all AI-assisted work. All AI assistance was carried out under the direction of the authors, and the resulting material was subsequently checked and verified by the authors. We take responsibility for the final content of this work, including text, claims, or artifacts produced with the aid of generative AI.
{
    \small
    \bibliographystyle{ieeenat_fullname}
    \bibliography{main}
}

\clearpage
\setcounter{page}{1}
\maketitlesupplementary

\section{Theory of Equivariant Replay}
\label{app:eer_theory}
We prove Proposition~\ref{prop:main_es_calibration} and , then give the an auxiliary result
for split-averaged inference.

\subsection{Proof of Proposition~\ref{prop:main_es_calibration}: ES Risk Calibration}
\label{app:eer_calibration}
Fix a noiseless past task $j$ and use the setup of
Sec.~\ref{sec:eer_risk_calibration}. Let
\[
W_M=\frac{M^\top M}{\rho_M}
    +\frac{(I-M)^\top(I-M)}{1-\rho_M},
\qquad 0<\rho_M<1,
\]
where $M$ is a binary diagonal mask. Set
$x_g=T_g^{-1}x$, $S_g=MA_jT_g$, $v_g=(My,S_g)$, and
$B_g=(A_jT_g)^\top W_M(A_jT_g)$.
By invariance, $x_g$ is independent of $(g,M)$ and has distribution $p_j$.
Since $My=S_gx_g$, conditioning on $v_g$ leaves $x_g$ and $(g,M)$
conditionally independent. Thus
$\mathbb E[x_g\mid v_g]=\mathcal F_j^\star(v_g)$ and
$\mathbb E[B_g\mid v_g]=Q_j(S_g)$.
The conditional quadratic decomposition of the ES risk gives
\begin{equation}
\mathcal R_j(\mathcal F)-\mathcal R_j^\star
=\mathbb E_{x,g,M}[d_g^\top Q_j(S_g)d_g],
\
d_g=\mathcal F(v_g)-\mathcal F_j^\star(v_g),
\label{eq:app_es_excess}
\end{equation}
with $\mathcal R_j^\star=\mathcal R_j(\mathcal F_j^\star)$.
Uniform group sampling, independence of $M$, and prior invariance give
\[
Q_j(ST_g)=T_g^\top Q_j(S)T_g,
\qquad
\mathcal F_j^\star(z,ST_g)=T_g^{-1}\mathcal F_j^\star(z,S).
\]
Together with reconstruction equivariance, these identities reduce
\eqref{eq:app_es_excess} to
\[
\mathcal R_j(\mathcal F)-\mathcal R_j^\star
=\mathbb E_{y,M}[d_M^\top Q_Md_M],
\]
where $Q_M=Q_j(MA_j)$ and
$d_M=\mathcal F(My,MA_j)-\mathcal F_j^\star(My,MA_j)$.
For fixed $y$, let
\[
\bar d=\bar Q_j^{-1}\mathbb E_M[Q_Md_M]
=\bar{\mathcal F}_j(y)-\bar{\mathcal F}_j^\star(y).
\]
Expanding the nonnegative quadratic form
$(d_M-\bar d)^\top Q_M(d_M-\bar d)$ conditionally on $y$, and using
$\mathbb E_M[Q_M\mid y]=\bar Q_j$, yields
\[
\bar\mu_j\mathcal D_j(\mathcal F)
\le\mathbb E_y[\bar d^\top\bar Q_j\bar d]
\le\mathbb E_{y,M}[d_M^\top Q_Md_M]
=\mathcal R_j(\mathcal F)-\mathcal R_j^\star,
\]
which proves the proposition.



\subsection{Split-Averaged Inference}
\label{app:eer_split_averaging}

Let $\widehat x(u)=T_g\mathcal F(My,MA_jT_g)$ for $u=(g,M)$ and
$\bar x=\mathbb E_u[\widehat x(u)\mid y,A_j]$.
For $S$ conditionally independent splits,
\begin{equation}
\mathbb E\left\|\frac1S\sum_{s=1}^S\widehat x(u_s)-x\right\|_2^2
=\mathbb E\|\bar x-x\|_2^2
+\frac1S\mathbb E\operatorname{tr}
\operatorname{Cov}(\widehat x(u)\mid y,A_j).
\label{eq:app_inference_variance}
\end{equation}
Increasing $S$ reduces the sampling variance of the reconstruction,
providing a direct computation--accuracy tradeoff for multi-split
inference. As in ES, this uniform average is a practical surrogate for
the coverage-weighted aggregate $\bar{\mathcal F}_j$ controlled in
Proposition~\ref{prop:main_es_calibration}.

\section{Additional Experiment Results}

\begin{figure}[t]
    \centering
    \includegraphics[width=0.95\linewidth]{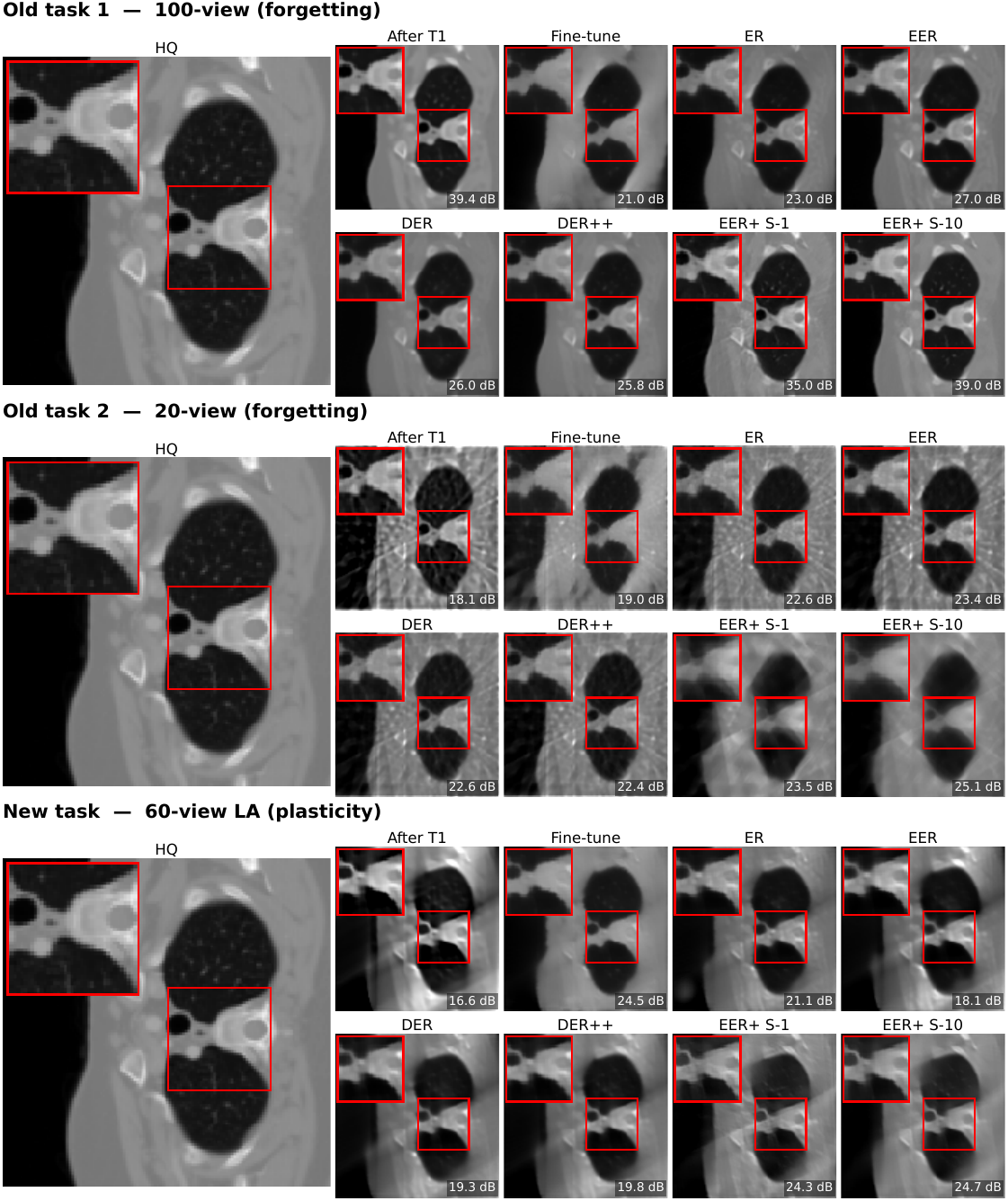}
    \caption{Visual results of sparse-view CT.}
    \label{fig:ct_qua_app}
\end{figure}

\begin{figure}
    \centering
    \includegraphics[width=0.95\linewidth]{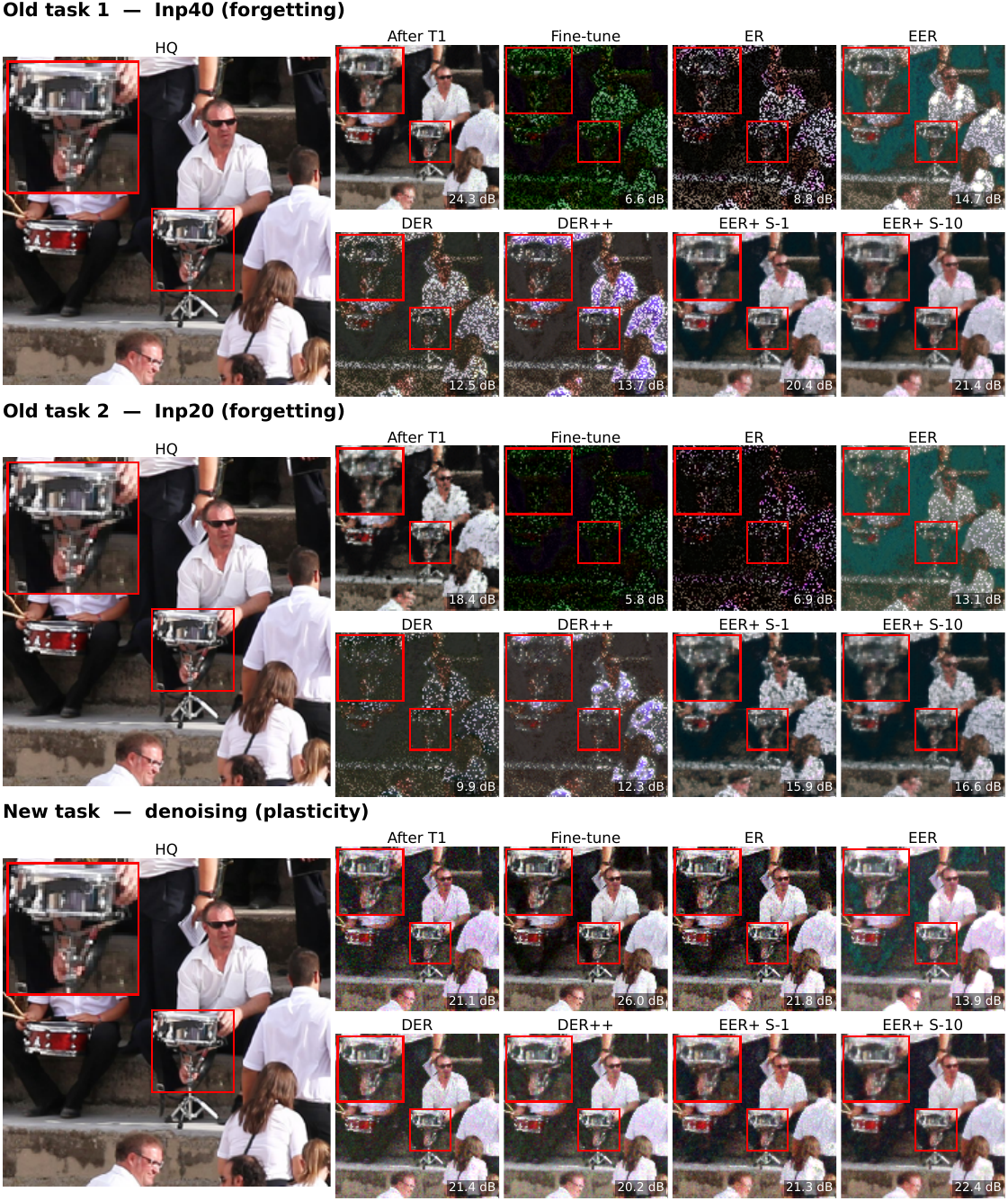}
    \caption{Visual results of natural imaging CL.}
    \label{fig:natur_qua_app}
\end{figure}

Figure~\ref{fig:ct_qua_app} and~\ref{fig:natur_qua_app} show extra visual comparison from Sec.~\ref{sec:exp_main}.

\section{Three-Task Ablation on Natural-Image Restoration}
\label{app:three_task_ablation}

We extend the replay-objective ablation to the three-task natural-image sequence $\mathrm{Inp40}\rightarrow\mathrm{Inp20}\rightarrow\mathrm{Denoise}$, training for 40 epochs at each subsequent stage. ES-only, ES+MC, and EER+ share the same current-task objective and differ only in replay supervision.

\begin{table}[t]
    \centering
    \caption{Equivariant replay improves retention across three restoration tasks. PSNR (dB) is evaluated after learning $\mathcal{T}_3$. S-1 and S-10 denote one and ten split reconstructions at inference. Best results within each inference setting are bold.}
    \label{tab:app_three_task_ablation}
    \small
    \setlength{\tabcolsep}{4pt}
    \renewcommand{\arraystretch}{1.08}
    \begin{tabular}{llcccc}
        \toprule
        Inf. & Method
        & Inp40 $\uparrow$
        & Inp20 $\uparrow$
        & Denoise $\uparrow$
        & Avg. $\uparrow$ \\
        \midrule
        S-10 & ES-only
        & 16.44 & 10.49 & \textbf{26.90} & 17.94 \\
        & ES+MC
        & 14.51 & 9.75 & 24.68 & 16.31 \\
        & EER+
        & \textbf{23.01} & \textbf{17.87}
        & 23.99 & \textbf{21.62} \\
        \midrule
        S-1 & ES-only
        & 15.52 & 10.18 & \textbf{25.49} & 17.06 \\
        & ES+MC
        & 13.67 & 9.45 & 22.54 & 15.22 \\
        & EER+
        & \textbf{22.10} & \textbf{17.28}
        & 22.48 & \textbf{20.62} \\
        \bottomrule
    \end{tabular}
\end{table}

EER+ achieves stronger retention on both past tasks while maintaining competitive new-task performance. With S-10 inference, it improves Inp40 and Inp20 PSNR over ES+MC by 8.50 and 8.12 dB, respectively, raising the three-task average from 16.31 to 21.62 dB. The advantage persists with S-1, where EER+ achieves comparable denoising accuracy (22.48 versus 22.54 dB) while substantially improving both past tasks. ES-only achieves the highest denoising PSNR but suffers greater forgetting, whereas ES+MC yields even lower past-task PSNR than ES-only. These results indicate that measurement consistency provides insufficient replay supervision for preserving unobserved image components.

\begin{table}[t]
    \centering
    \caption{EER+ reduces forgetting during the final task transition. Values denote PSNR decreases (dB) from each method's $\mathcal{T}_2$ checkpoint to $\mathcal{T}_3$, evaluated with S-10 inference. Lower is better.}
    \label{tab:app_three_task_stage_forgetting}
    \small
    \setlength{\tabcolsep}{8pt}
    \renewcommand{\arraystretch}{1.08}
    \begin{tabular}{lcc}
        \toprule
        Method
        & Inp40 $\downarrow$
        & Inp20 $\downarrow$ \\
        \midrule
        ES-only & 6.31 & 10.92 \\
        ES+MC   & 5.49 & 7.26 \\
        EER+    & \textbf{1.54} & \textbf{4.24} \\
        \bottomrule
    \end{tabular}
\end{table}

EER+ also limits forgetting during the final transition, reducing the PSNR losses on Inp40 and Inp20 by 3.95 and 3.02 dB relative to ES+MC (Table~\ref{tab:app_three_task_stage_forgetting}). This confirms that its retention advantage extends beyond the performance established at $\mathcal{T}_2$. Together, these results demonstrate that equivariant splitting provides more effective replay supervision across successive operator shifts.

\end{document}